\RequirePackage{lineno}
\documentclass[superscriptaddress,prd,twocolumn,showpacs,preprintnumbers,amsmath,amssymb,bibnotes,secnumarabic,altaffilletter,floatfix,aps,nofootinbib,10pt]{revtex4-1}

\usepackage{epsfig}
\usepackage{bigstrut}
\usepackage{amsmath}
\usepackage{graphicx}
\usepackage{epsfig}
\usepackage{fancyhdr}
\usepackage{calc} 
\usepackage{float}
\usepackage{afterpage}
\usepackage{lineno}
\usepackage{natbib}
\usepackage[usenames]{color}

\linenumberdisplaymath
\usepackage[colorlinks=true, breaklinks=true,pdfstartview=FitV, linkcolor=black, citecolor=black, urlcolor=black]{hyperref}

\begin{document}

\title{A potential sterile neutrino search utilizing spectral distortion in a two-reactor/one-detector configuration}

\author{M. Bergevin}
\author{C. Grant}
\author{R. Svoboda}
\address{Department of Physics, University of California, Davis, One Shields Ave, Davis, CA 95616}

\date{\today}

\begin{abstract}
There is an observed deficit of about 6\% in the expected rate of anti-neutrino interactions when averaging over many different reactor experiments. While the significance of the deficit is low (98.6 \% CL), there is speculation that a non-interacting ``sterile" neutrino could be the cause. In this paper we explore the possibility of a two-reactor/one-detector experiment at intermediate distances (100-500 meters) to look for a sterile neutrino in the mass range implied by this deficit.  A method for probing $\Delta m^2$ phase space is developed using interference patterns between two oscillated spectra at different baselines. This method is used to investigate the potential sensitivity of the Double Chooz experiment, which has a single Near Detector at distances of 351 m and 465 m from two reactors of identical design. We conclude that Double Chooz could investigate sterile neutrino in the $\Delta m^{2}$ range of 0.002 to 0.5 eV$^2$ over 5 years of near detector running. 
\end{abstract}

\maketitle
\section{Introduction}
There are global hints of the possible existence of sterile neutrinos from accelerator experiments~\cite{ACCEXP, ACCEXP1, ACCEXP2, ACCEXP3}, reactor experiments~\cite{REACTEXP} and cosmological measurements~\cite{COSMEXP,COSMEXP1}. While the existence of one or more sterile neutrinos is not the only possible explanation for these results, it nevertheless becomes interesting to devise new experiments that explore this possibility. In this paper we investigate the sensitivity of the Double Chooz experiment to detect the oscillations of electron flavor neutrinos into sterile neutrinos, and propose a new experiment optimized for this search.

There have been nearly 20 reactor neutrino experiments at distance of 10-100 meters from reactor cores.
In a recent review paper~\cite{REACTEXP} the expected rates were re-calculated using up-to-date reactor anti-neutrino flux predictions and a 3-flavor neutrino oscillation hypothesis. The authors found that on average there is a $\sim$6\% deficit in the rate of observed anti-neutrino interactions measured to that expected. This has come to be called the Reactor Anti-neutrino Anomaly (RAA), and has been taken as an indication that the three-flavor oscillation hypothesis may not be complete. To follow up this hypothesis, the authors performed global fits using rate information from these experiments assuming a three active flavor neutrino and one sterile neutrino oscillation model (also known as a 3+1 model).  Folding the results of these fits in with spectral shape constraints from the Bugey-3 experiment~\cite{Bugey}, they calculated best fit oscillation parameter values of $ \Delta m^{2}_{14} = 1.5$ eV$^2$ and $\sin^2(2\theta_{14})=0.14$~\cite{REACTEXP}.

In this paper we present a new method of gaining increased sensitivity to 3+1 oscillation parameters. This method utilizes two reactors and one detector, referred to as the "two-reactor/one-detector" configuration, and relies on the differences in the measured spectrum from running each reactor singly.  Thus, this method is effective only in the case of two reactor cores, each of which runs a significant period of time while the other core is down. This is indeed the case for the Chooz B reactor configuration and the Double Chooz near detector. The near detector will be 351 m  and 465 m from two 4.25 GW$_{\mathrm{th}}$ power reactors. With the analysis technique described here, this configuration will lead to sensitivity in a region of $\Delta m^{2}_{14}$ and $\sin^2(2\theta_{14})$ that has not been previously explored.

\section{Shape Analysis of Sterile Neutrinos}\label{ShapeAnal}
To help give clarity to the discussion, a simple example is described using the two-neutrino oscillation formalism.  The two-neutrino survival probability can be described by the following formula:
\begin{eqnarray}  
P_{ee}& =&  1 -\sin^2(2\theta)\sin^2\left( \frac{ \Delta m^{2} L }{4E_{\bar{\nu}_e}} \right)\nonumber\\
&=& 1- \alpha^2\sin^2\left(\beta L\right)
 \label{sterSurv} 
\end{eqnarray}
where $\alpha\equiv\sin(2\theta)$ and $\beta(E_{\bar{\nu}_e}) \equiv { \Delta m^{2}}/{4E_{\bar{\nu}_e}} $.  For a single detector measuring the disappearance of $\bar{\nu}_e$'s from a single reactor, the sensitivity to $\Delta m^2$ and $\sin^2(2\theta)$ depends on the reactor-detector distance, $L$.  For an appropriate $L$ the visible $E_{\bar{\nu}_e}$ spectrum will be distorted due to the $L/E_{\bar{\nu}_e}$ dependence of the survival probability.  It is possible to visually enhance this spectral distortion by forming a ratio using an independent $E_{\bar{\nu}_e}$ spectrum obtained at a different distance where sensitivity to $\Delta m^2$ and $\sin^2(2\theta)$ is also expected.  In this paper we present a mathematical formalism for a two-reactor/one-detector configuration. It is expected that many (but not all) detector systematics will cancel if the single-reactor running periods are roughly equally interspersed. This is discussed in more detail in section \ref{exampleDC} . In addition, although the detector backgrounds are not expected to vary with reactor running period, we assume for the formalism that the relevant backgrounds have been subtracted.  We note that this work was initially motivated to explain the tension between the rate and spectral measurements found in the $\theta_{13}$ analyses of the Double Chooz, Daya Bay and Reno experiments, which is yet not fully understood~\cite{DCFar,DCFarH,RENO,DayaBay}. 

There are four distances (three of which are independent) that can be defined for the two-reactor/one-detector configuration:
\begin{eqnarray}
(a) &&L_1 \equiv \mbox{distance from detector to reactor } R_1\nonumber\\
(b) &&L_2 \equiv \mbox{distance from detector to reactor } R_2\nonumber\\
(c)&&L_{2-1}\equiv L_2 - L_1\nonumber\\
(d)&&L_{1+2} \equiv L_1 + L_2 \nonumber
\label{baseline}
\end{eqnarray}
 
\noindent Using the form of Equation \ref{sterSurv}, the ratio of the energy spectra obtained from the reactors $R_1$ and $R_2$ can be written as:

\begin{equation}
\frac{P^{R_1}_{ee}}{P^{R_2}_{ee}} = 
\frac{1 -\alpha^2\sin^2(\beta L_{1})}{1 -\alpha^2\sin^2(\beta L_{2})}.
\label{simpleRatio}
\end{equation}

\noindent This ratio can be expressed in a more useful form by multiplying the numerator and denominator by a factor of $1 + \alpha^2\sin^2(\beta L_2)$.  After simplifying, the expression in equation \ref{simpleRatio} becomes:

\begin{equation}  
\begin{aligned}
& \frac{P^{R_1}_{ee}}{P^{R_2}_{ee}} =  \\
& \frac{1+\alpha^2\sin\left(\beta L_{2-1} \right)\sin\left(\beta L_{1+2} \right)-\alpha^4\sin^2(\beta L_1)\sin^2(\beta L_2)}{1-\alpha^4\sin^4(\beta L_2)}\\
\end{aligned}
\label{mixedRatio}
\end{equation}
\vspace{1mm}

\noindent where the spectral distortions due to the mixtures of $L_1$ and $L_2$ in sine functions are clearly revealed.   The $\alpha^2\sin\left(\beta L_{2-1} \right)\sin\left(\beta L_{1+2} \right)$ creates modulations on the spectral shape at both higher and lower frequency than the modulation produced by baselines $L_1$ or $L_2$ alone.  In the limit that $\alpha$ is small, Equation \ref{mixedRatio} may be expressed as:

\begin{equation}
\begin{aligned}  
&\frac{P^{R_1}_{ee}}{P^{R_2}_{ee}} \approx  \\
&1 + \left[1 -\alpha^2\sin^2(\beta L_2)\right] \left[ \alpha^2\sin\left(\beta L_{2-1} \right)\sin\left(\beta L_{1+2} \right)\right] \\
& + O(6) + ...\\
\end{aligned}
\end{equation}

\noindent Here, the ratio is a function of the antineutrino survival probability at a baseline of $L_2$, convoluted with modulations of the spectral shape at baselines $L_{1-2}$ and $L_{1+2}$. The terms $L_{1-2}$ and $L_{1+2}$ represent different frequencies in L/E space, which give rise to a spectral distortion of the antineutrino survival probability in the visible energy range of $1-9$ MeV.  Of course, complication arises when including detector energy resolution, which will reduce the contrast in high frequency features of the ratio of survival probabilities.  These detector effects will be covered in the next section, where we investigate the sensitivity of this technique using the Double Chooz near detector. 

It is worth pointing out that this method also applies to the case where the oscillation probabilities are either added or subtracted:

\begin{eqnarray}
{P^{R_1}_{ee}}+{P^{R_2}_{ee}} &=& (1 -\alpha^2\sin^2(\beta L_{1})) + (1 -\alpha^2\sin^2(\beta L_{2})) \nonumber\\
 &=& 2 - \left(\alpha^{2} - \alpha^{2}\frac{\cos(2\beta L_{1}) + \cos(2\beta L_{2})}{2} \right)\nonumber\\
 &=& (2 - \alpha^{2}) + \alpha^{2}\cos\left(\beta L_{1+2} \right)\cos\left(\beta L_{1+2} \right)\nonumber\\
 & &
\end{eqnarray}

\begin{eqnarray}
{P^{R_1}_{ee}}-{P^{R_2}_{ee}} &=& (1 -\alpha^2\sin^2(\beta L_{1})) - (1 -\alpha^2\sin^2(\beta L_{2}))\nonumber\\
 &=&\alpha^2\sin\left(\beta L_{2-1} \right)\sin\left(\beta L_{1+2} \right)\nonumber\\
 & &
\end{eqnarray}

\section{Double Chooz near detector sensitivity}
\label{exampleDC}The near detector is positioned at $L_1 = 351$ m and $L_2 = 465$ m from Chooz B reactors $R_1$ and $R_2$.  A five year run time for this detector (assuming a down cycle of 15\% per reactor, which was the case for their recent publication \cite{DCFarH}) will yield 548 days of data taken with only either $R_1$ or $R_2$ on.  Each reactor operates at 4.25 GW$_{\mathrm{th}}$, giving expected rates of about 230 $\bar{\nu}_e$'s per day from $R_1$ and 130 $\bar{\nu}_e$'s  per day from $R_2$. 

For comparison sake, the Double Chooz far detector, positioned at $L_1 = 998$ m and $L_2 = 1115$ m, has been taking data since April 2011 \cite{DCFar,DCFarH}. The measured rates are about 28 $\bar{\nu}_e$'s per day from $R_1$ and 23 $\bar{\nu}_e$'s per day from $R_2$.  The technique developed in the previous section is statistically limited due to the $1/L^{2}$ fall off of neutrino intensity; as such, only the near detector is suitable for such a study.

The near detector can offer sensitivity to sterile neutrinos not only by making a shape measurement at $L_{1}$ and $L_{2}$, but also by identifying the interference terms $L_{2-1}$ and $L_{1+2}$.  There exists a frequency modulation at a distance of $L_{2-1}= 114$ m, a region which has already been probed by previous reactor experiments located at a distance of $L \sim 100$ m from the source.   
The near detector is not optimized for probing large $\Delta m^2$ regions of the sterile neutrino phase space, but will have sensitivity for values of $\Delta m^2$ less than $10^{-2}$ eV$^{2}$.  The Bugey-3 spectral shape analysis was not sensitive to this region.

\begin{table}[bbbb]
{
\footnotesize
\caption[]{Distance from Reactor R$_1$ and R$_2$ to the center of the detector from geodesic surveys.}
\begin{tabular}{l l l l l } 
\hline
& L$_{2-1}$&  L$_{1}$ & L$_{2}$ &L$_{1+2}$  \\
\hline
Near (m)                   &     114  &351  &  465 & 816 \\
Far  (m)                     &    117    & 998 & 1115  & 2113 \\
\hline
\end{tabular}
\label{NearFarProbe}
}
\end{table}

The formalism developed in Section \ref{ShapeAnal} is applicable when there are no backgrounds present. In actual experiments, backgrounds such as $^{9}$Li and fast neutrons will have small contributions which must be taken into account.  By performing a two-reactor/one-detector study of the backgrounds, complimented by the availability of reactor off data when there are only two cores, we assume a good understanding of these quantities can be obtained.  In turn, these backgrounds may then be subtracted from the far and near reactor data with a small loss of sensitivity compared to statistical and detector resolution uncertainties.

\begin{figure}[t]
 \centering
\includegraphics[width=0.49\textwidth]{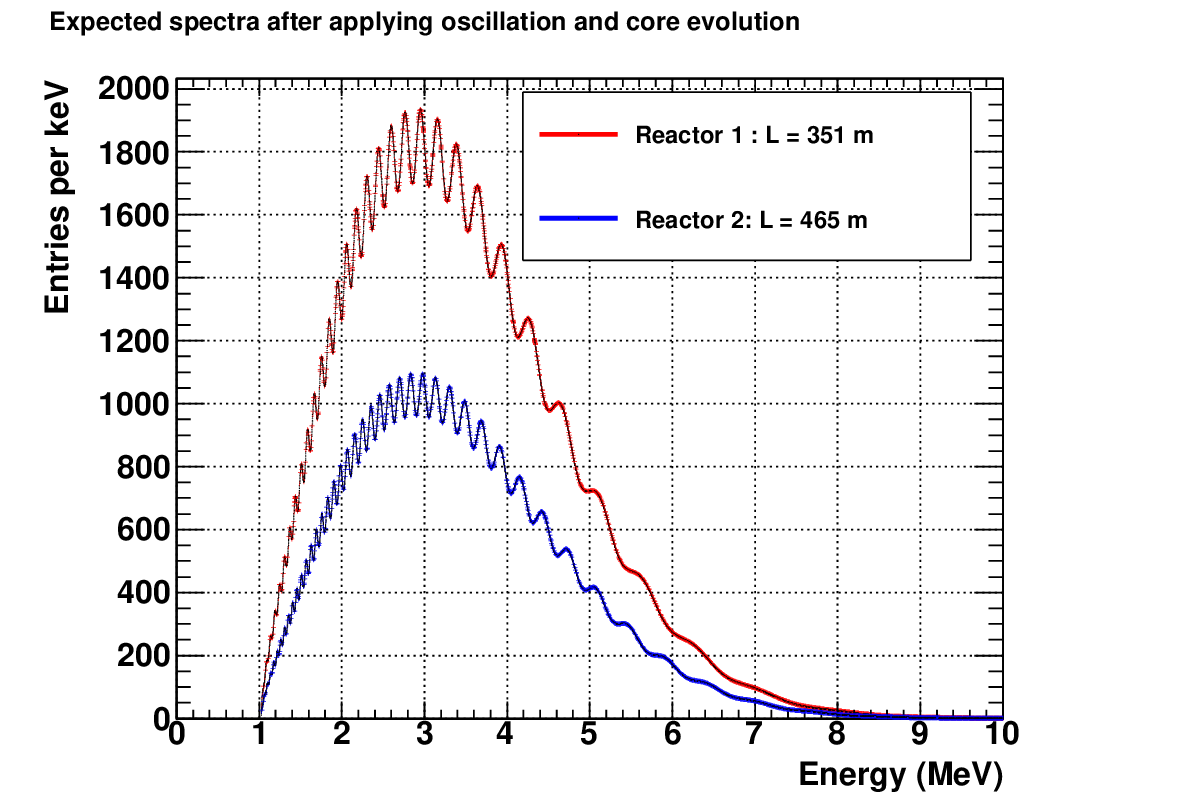}
\caption[]{Expected $R_1$ and $R_2$ reactor spectra for the Double Chooz near detector without detector energy resolution assuming $\Delta m_{new}^{2} = 0.5$ eV$^2$ and $\sin^{2}(2\theta_{new})=0.12$.}
\label{spectralSHAPE}
\end{figure}

Figure \ref{spectralSHAPE} shows the $R_1$ and $R_2$ visible energy spectra for values of $\Delta m_{14}^{2}$=0.5 eV$^2$ prior to adding energy resolution.  The following systematic uncertainties on the observed spectral shapes have been incorporated in the ratio analysis of $R_1$ and $R_2$ data:
\begin{itemize}
 \item  Energy resolution:  (7 $\pm$ 1)\%/$\sqrt{\mathrm{E[MeV]}}$ from Double Chooz far detector).
 \item  Detector stability:   estimated at 1\% from the Double Chooz far detector \cite{DCFar} measurements of the stability of the Gd capture gamma peak.
 \item Reactor core size:   3.47 m diameter for the Chooz B reactors. We have assumed the neutrinos start randomly inside this core.
 \item  Fuel loading:   contributes $<$ 0.01\% uncertainty, based on studies where extremes of fuel loading uncertainties were used as inputs.
 \end{itemize}

\begin{figure}[t]
 \centering
\includegraphics[width=0.484\textwidth]{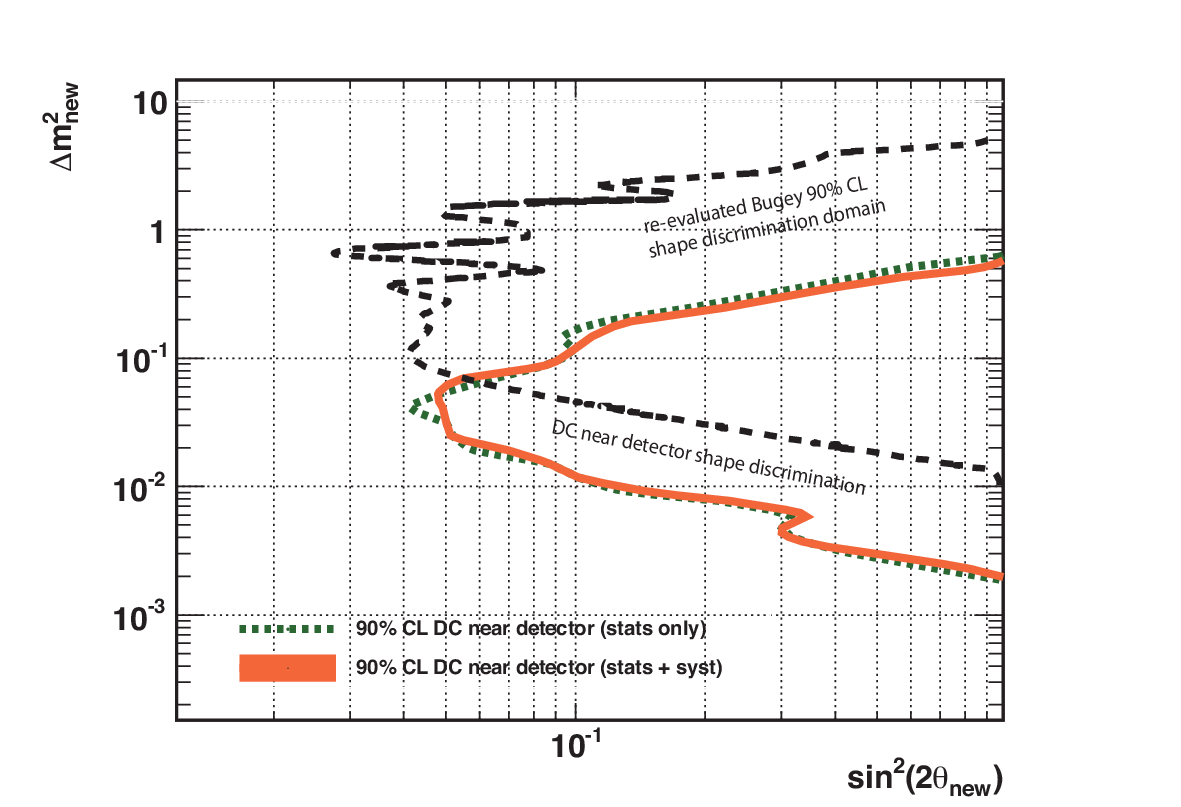}\caption[]{Exclusion domain of $\Delta m^{2}_{new}$ for the case of the Double Chooz near detector. }
\label{rasterScan}
\end{figure}

 \vspace{0.05in}
The combined effect of these uncertainties to the sensitivity of sterile neutrino oscillations are shown in Figure \ref{rasterScan}. In fact, they have very little impact on the exclusion domain of this technique with the near detector after 5 years of detector operation. The analysis is clearly dominated by statistical uncertainty. The spectra from the re-evaluation of the Bugey-3 shape discrimination \cite{REACTEXP} is also included as a comparison.  

\section{Discusion}
It should be noted that the mathematical formalism developed in section \ref{ShapeAnal} is generic and can be applied to any two-reactor/one-detector experiment.  The positive discovery of sterile neutrinos will require strong evidence.  A strong shape discrimination and good statistics are paramount to any new proposed experiments. 

There are a few guidelines required to optimize this technique for new experiments.  First and foremost, the distances between the detector and the far and near reactors cannot be equal since the interference term would vanish.   The strength of shape discrimination is correlated with the interference terms $L_{2-1}$ and $L_{2+1}$.  One can envision an experiment with $L_{2-1}=10\sim15$ m that can devised such that the ILL region can be probed, however, the reactor core size will add greater uncertainty at short distances and will have to be modeled carefully.  

New experiments must also seek to improve spectral shape sensitivity.  The principle culprit for the loss of shape discrimination power is the energy resolution of a detector.  Ratio analyses for different values of $\Delta m^{2}_{14}$ are shown with no smearing applied in Figure \ref{noSmearing} and with smearing applied in Figure \ref{smearing}.  If sensitivity to larger $\Delta m^{2}_{14}$ is desired, detectors with better energy resolution is a vital requirement.

Currently, there are few experiments which have sensitivity to $\Delta m^{2}_{new}$ below the limits set by Bugey-3.  The ICARUS experiment can claim sensitivity in this region, however it is a $\nu_\mu\rightarrow\nu_e$ appearance experiment and will not address the probability of $\bar{\nu}_e$ disappearance. If a solution to the RAA exists in the form of sterile neutrino oscillations, such a solution could be probed by reactor $\bar{\nu}_e$ experiments that cleverly address the many unknowns present in reactor $\bar{\nu}_e$ flux predictions.  

The most recent results from Planck report $N_{eff} = 3.06$  \cite{COSMEXP1}.  When combined with BAO, $H_{0}$ data, and other CMB measurements, $N_{eff} = 3.30\pm0.27$. This mildly disfavors four flavors of neutrinos but does not completely rule out this possibility. If the sum of all neutrino masses were due to a sterile flavor, then the region below $(0.23)^{2} \sim 0.053$ eV$^{2}$ becomes even more interesting to explore with this new technique.
\afterpage{%
\begin{figure}[H]
 \centering
\includegraphics[width=0.46\textwidth]{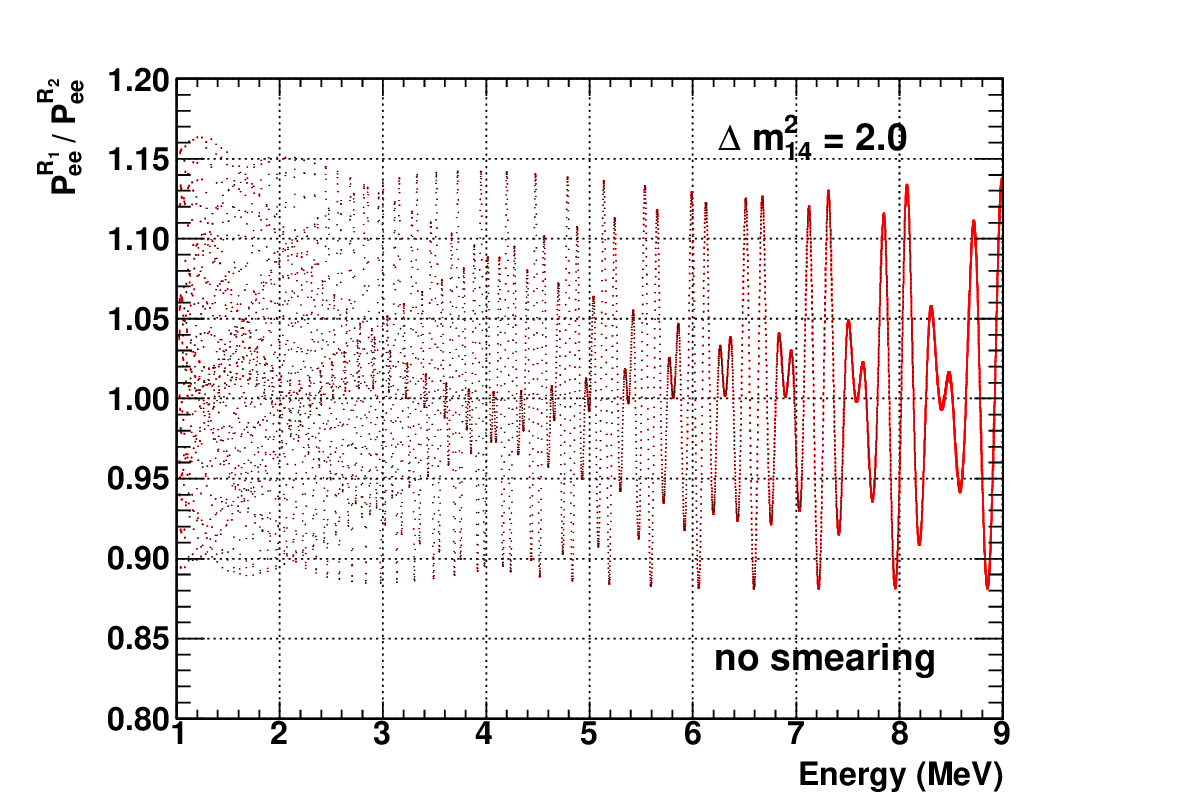}
\includegraphics[width=0.46\textwidth]{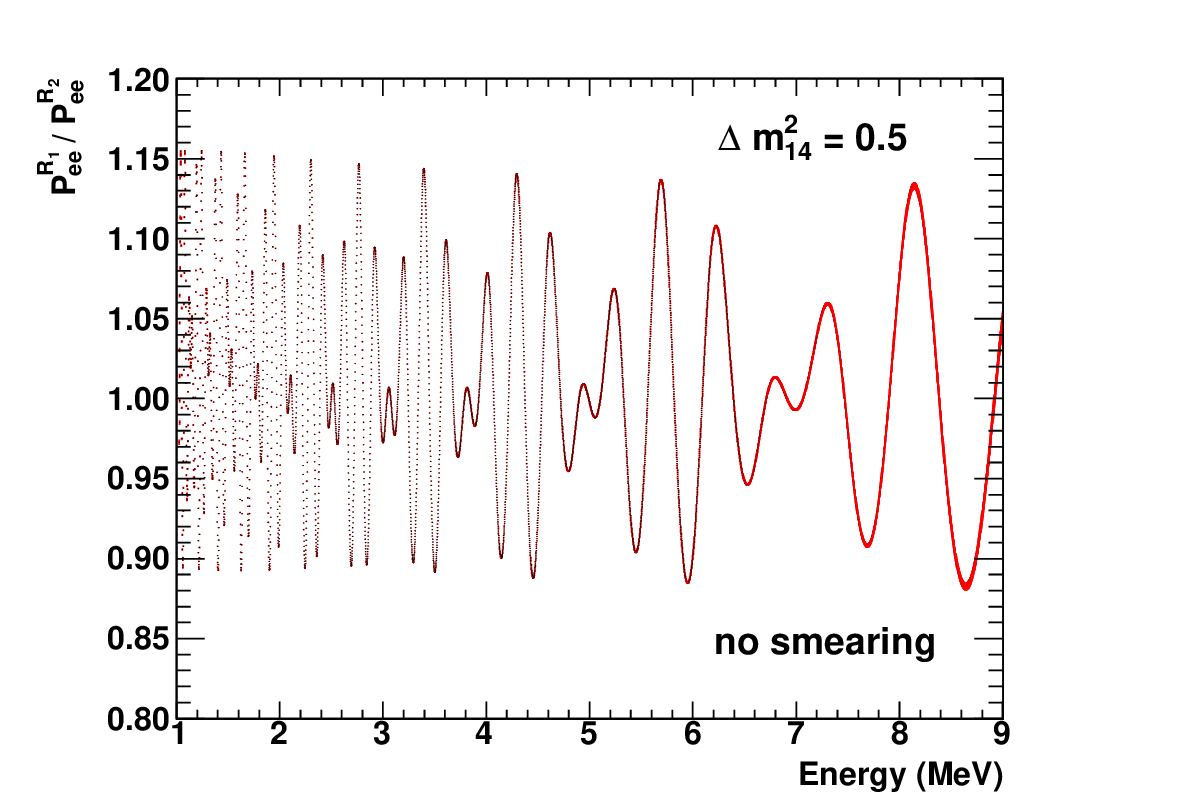}
 \includegraphics[width=0.46\textwidth]{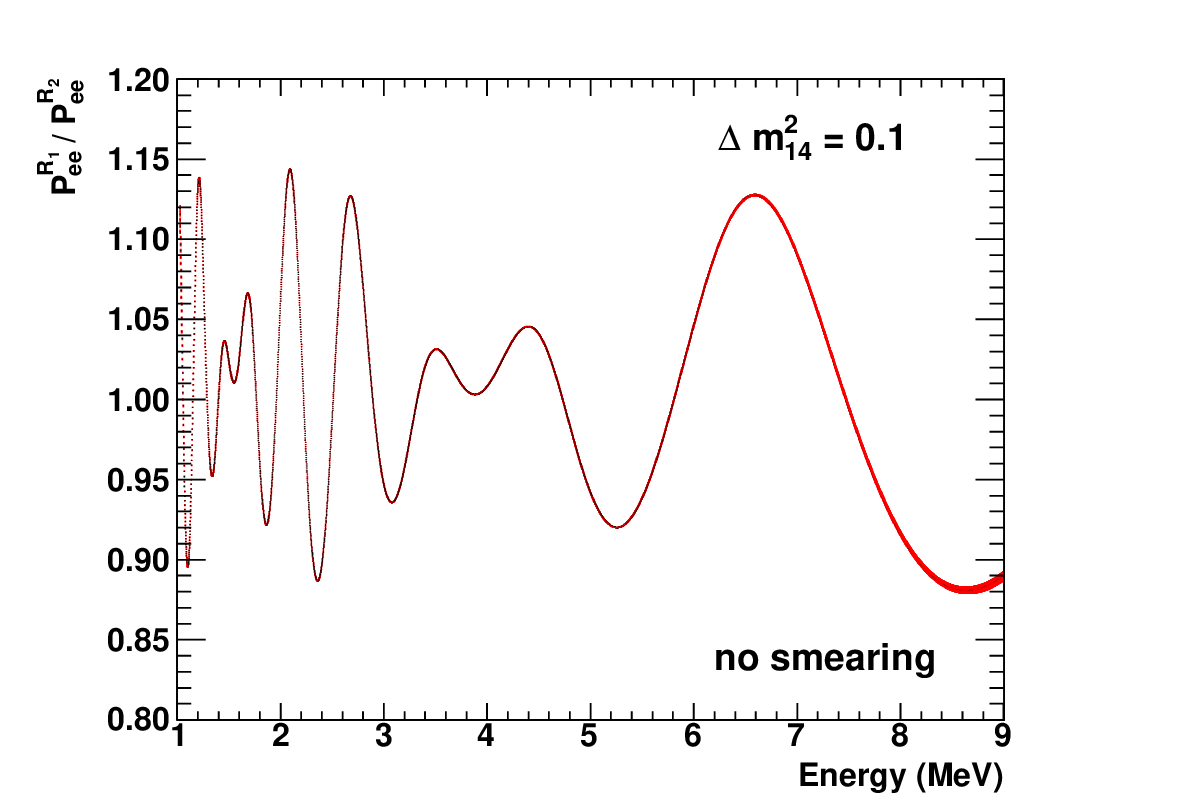}
  \includegraphics[width=0.46\textwidth]{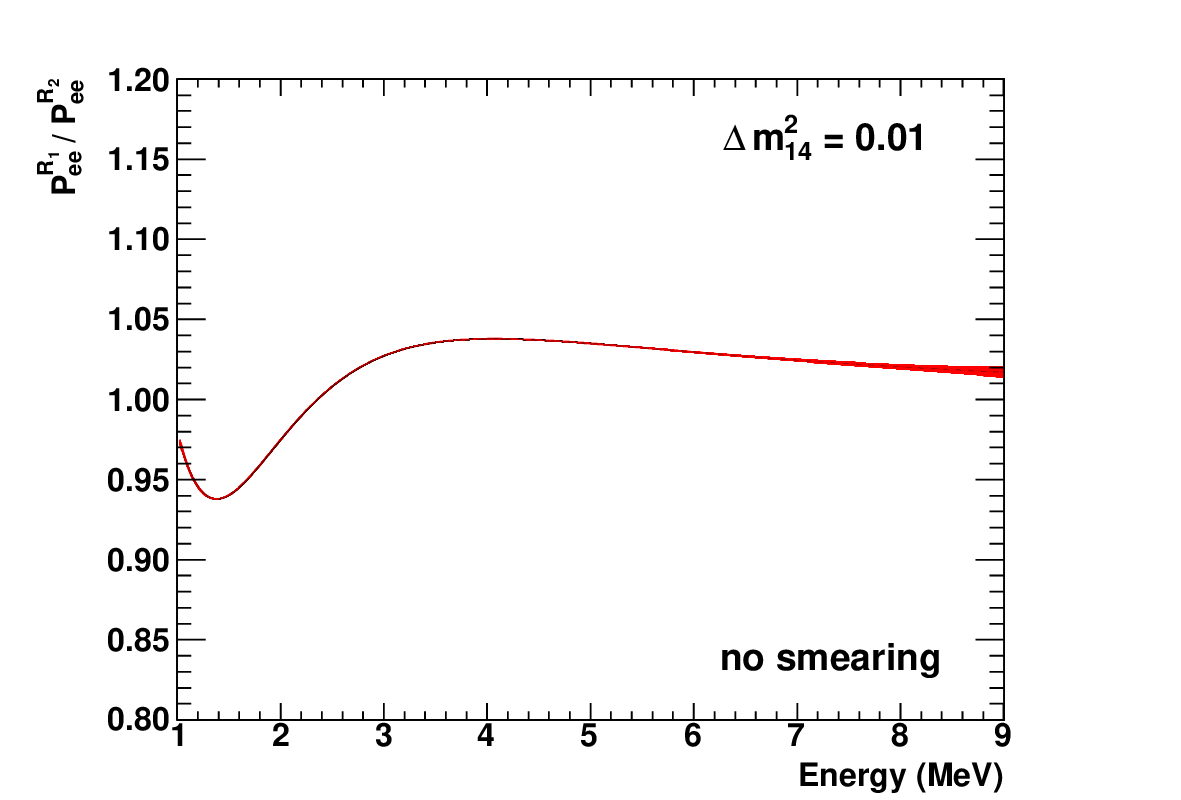}
\caption[]{Ratio analysis for value of $\Delta m^{2}_{14}$ respectively of 2.0, 0.5, 0.1 and 0.01 eV$^2$ without detector energy resolution. } 
\label{noSmearing}
\end{figure}

\begin{figure}[H]
 \centering
 \includegraphics[width=0.46\textwidth]{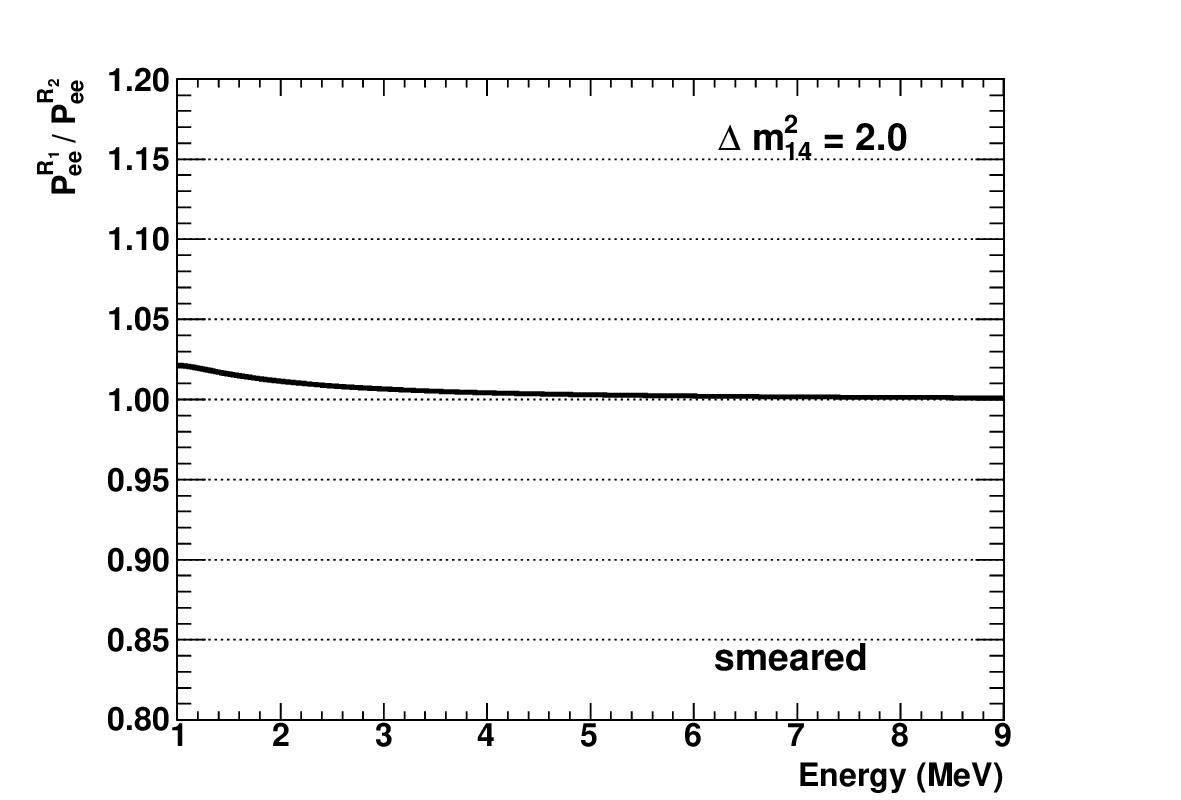}
\includegraphics[width=0.46\textwidth]{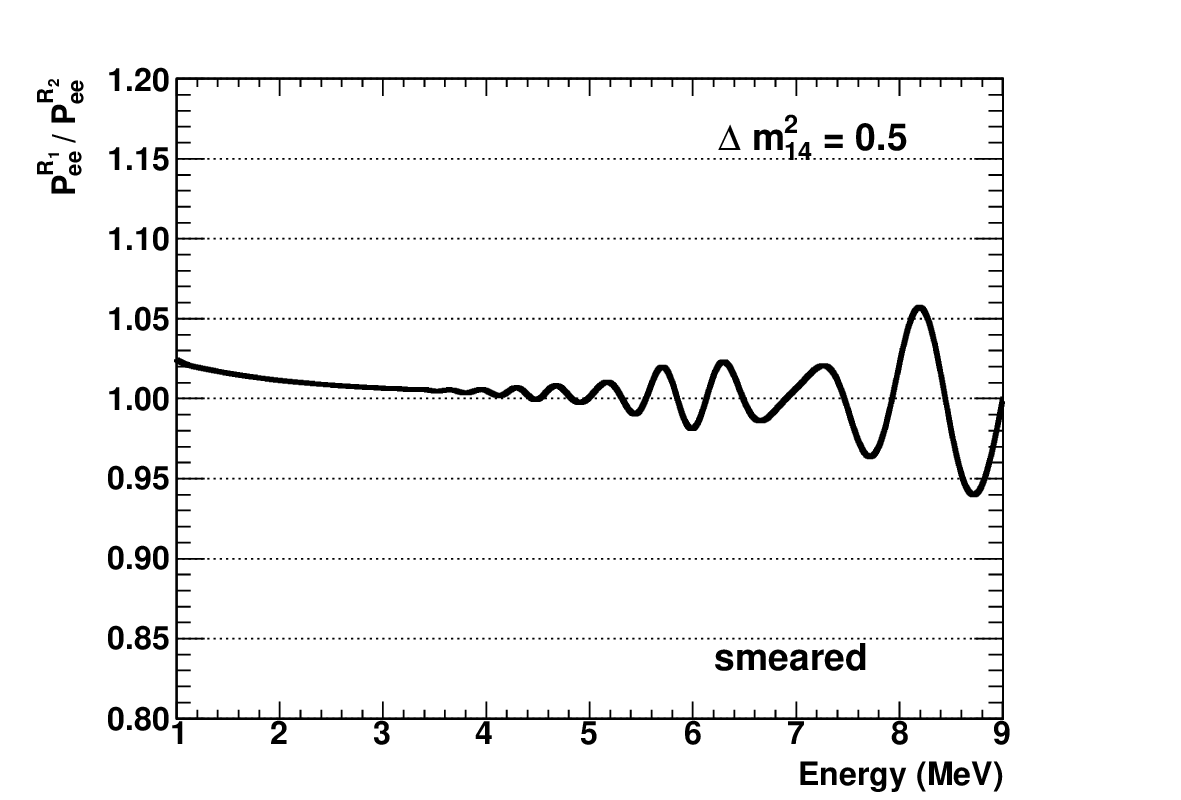}
\includegraphics[width=0.46\textwidth]{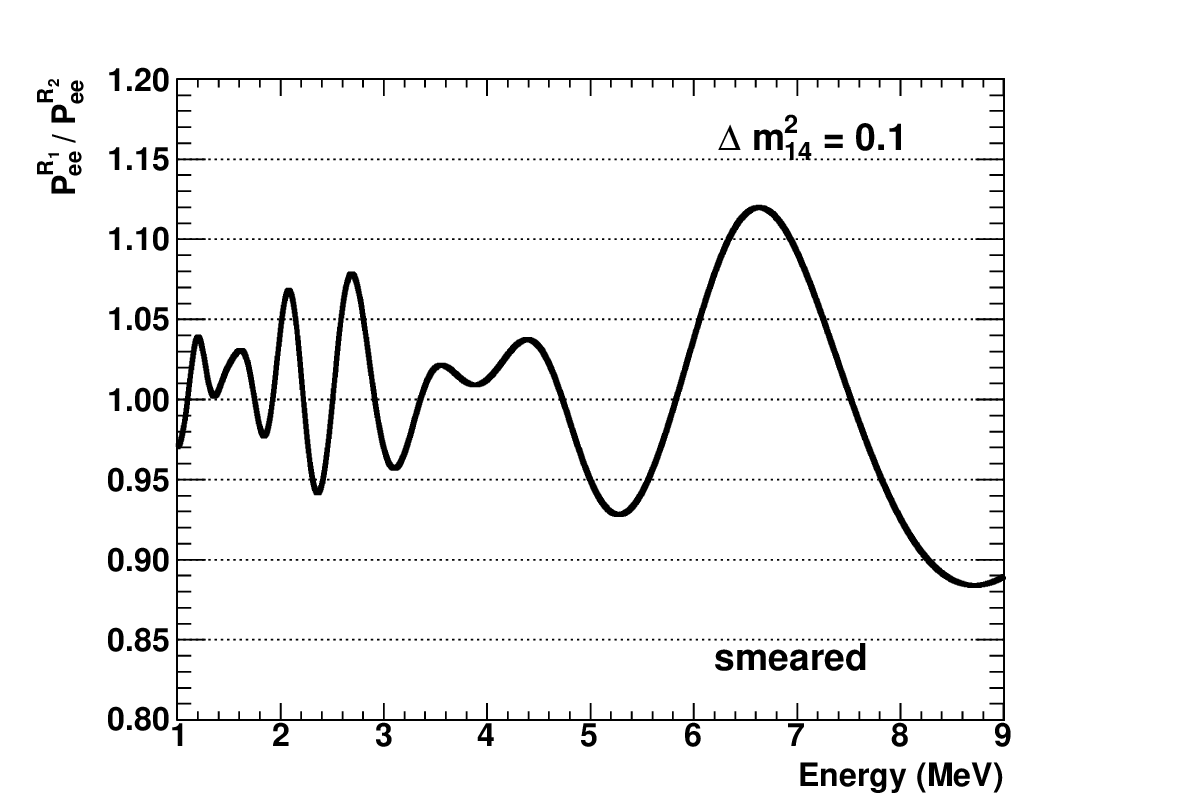}
 \includegraphics[width=0.46\textwidth]{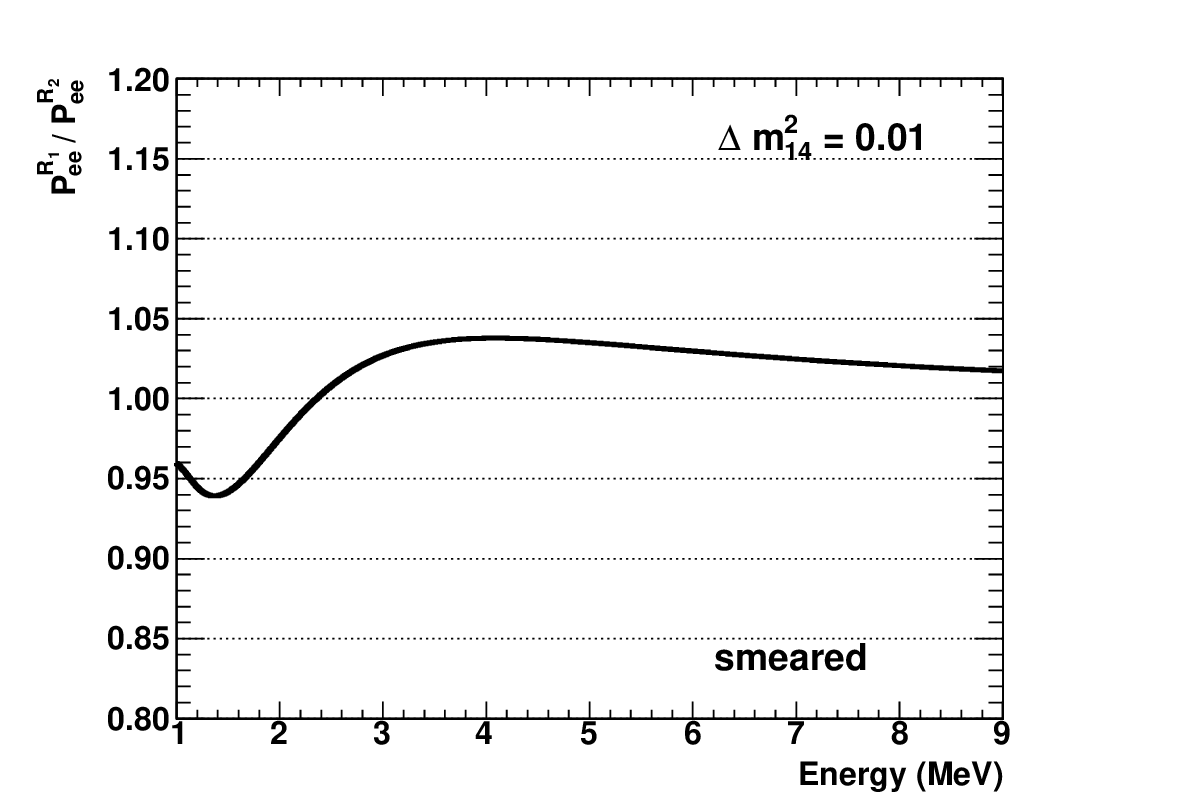}
 \caption[]{Effect of adding 7\%$/\sqrt{\mathrm{E[MeV]}}$ detector resolution on the ratio analysis for $\Delta m^{2}_{14}$ values of 2.0, 0.5, 0.1 and 0.01 eV$^2$.}  
 \label{smearing}
\end{figure}
\clearpage
}

\section{Conclusion}\label{sec:Conclusion}
The two-reactor/one-detector technique presented in this paper provides a formalism that could be used to study sterile neutrino oscillations as a viable solution to the RAA.  The technique utilizes shape distortions in anti-neutrino spectra, which are caused by interference terms coming from the addition, subtraction, or division of oscillation probabilities over different source-detector baselines.  The shape discrimination power relies on the backgrounds being well-understood, knowledge that can be obtained when data is taken during periods when both reactors are off.  In the case of Double Chooz, the distances between the Chooz B reactors and the near detector will provide sensitivity to a new region of $\Delta m^{2}_{14}$ from 0.002 to 0.5 eV$^2$.  Future experiments could seek to optimize the distances involved in a two-reactor/one-detector analysis to increase spectral shape discrimination in the $\Delta m^{2}$ region of interest. 

\bibliography{ratio_biblio}

\end{document}